\documentclass[preprint,showpacs,preprintnumbers,showkeys,amsmath,amssymb]{revtex4}

\usepackage{graphicx} % Include figure files
\usepackage{dcolumn}  % Align table columns on decimal point
\usepackage{bm}       % bold math
\usepackage{amssymb}

\begin{document}

%\preprint{APS/123-QED}

\title{Avalanches and Generalized Memory Associativity 
       in a Network Model for Conscious and Unconscious Mental Functioning}
% Force line breaks with \\

\author{Maheen Siddiqui}
  \email{Maheen.faisal.14@ucl.ac.uk}
  \affiliation{Centre for Brain and Cognitive Development, 
  Birkbeck College, University of London,
  London WC1E 7HX, United Kingdom.}%

%\author{Maheen Siddiqui}%
%  \email{maheen.siddiqui12@imperial.ac.uk }
%  \affiliation{Department of Mathematics, 
%  Imperial College London, South Kensington Campus
%  Electrical Engineering Building - EEE1201
%  London SW7 2AZ, United Kingdom.}%

\author{Roseli S. Wedemann, Corresponding Author }%
  \email{roseli@ime.uerj.br}
  \affiliation{Instituto de Matem\'atica e Estat\'\i stica,   
  Universidade do Estado do Rio de Janeiro,
  Rua S\~ao Francisco Xavier 524, 20550-013, Rio de Janeiro, Brazil.}%

\author{Henrik Jensen}%
  \email{h.jensen@imperial.ac.uk}
  \affiliation{Centre for Complexity Science and Department of Mathematics, 
  Imperial College London,
  London SW7 2AZ, United Kingdom,\\
  and\\
  Institute of Innovative Research, Tokyo Institute of Technology, 
  4259, Nagatsuta-cho, Yokohama 226-8502, Japan.}%

\date{\today}% It is always \today, today,

\begin{abstract}

We explore statistical characteristics of avalanches associated 
with the dynamics of a complex-network model, where two modules corresponding to 
sensorial and symbolic memories interact, 
representing unconscious and conscious mental processes.
The model illustrates Freud's ideas regarding the neuroses and that consciousness 
is related with symbolic and linguistic 
memory activity in the brain. It incorporates the Stariolo-Tsallis generalization
of the Boltzmann Machine in order to model memory retrieval and associativity.
In the present work, we define and measure avalanche size distributions
during memory retrieval, in order to gain insight
regarding basic aspects of the functioning of these complex networks.
The avalanche sizes defined for our model should be related to the 
time consumed and also to the size of the neuronal 
region which is activated, during memory retrieval.
This allows the qualitative comparison of the behaviour of the 
distribution of cluster sizes, obtained during fMRI measurements of the
propagation of signals in the brain, 
with the distribution of avalanche sizes 
obtained in our simulation experiments. This comparison
corroborates the indication that the Nonextensive Statistical 
Mechanics formalism may indeed be more well suited to model the 
complex networks which constitute brain and mental structure.

\end{abstract}

%\pacs{87.19.ll, 87.19.lv, 89.75.Fb}% PACS, the Physics and Astronomy
%Use showkeys class option if keyword

\maketitle

\section{\label{sec:Introd} Introduction}

The purpose of the present work is to define, measure and study 
statistical properties of avalanches which occur during memory 
retrieval in an artificial neural network, developed to model
neurotic phenomena and the associated conscious/unconscious 
interactions in mental life~\cite{We09}. 
It was famously reported by Freud~\cite{Fr74,Fr53} that
patients with neurotic symptoms systematically repeated 
these symptoms in the guise of ideas and impulses.
This tendency was referred to as the \textit{compulsion to repeat\/}.
The patients' repetitive behaviour suggested to Freud that
they had ``\textit{\ldots the intention of correcting
a distressing portion of the past\/}\ldots'' \cite{Fr66}.
Although the patient is aware of the obsessional ideas 
and impulses, and of the performance of the neurotic actions, 
the concomitant psychical predeterminants remain unconscious. 
In order to infer these predeterminants, and to bring them under the light 
of consciousness, Freud developed the analytical treatment
called \textit{working-through\/}. This treatment provides,  
through the interpretation of the predeterminants by freely talking 
in psychoanalytical sessions, the connections into which they are inserted.

One of the early seminal findings of psychoanalytic research is that neurotic symptoms
arise from traumatic and repressed memories. In fact, the repressed material can 
be construed as knowledge that, in spite of being present in the
subject, is inaccessible to him through symbolical representation. 
In other words, this knowledge is momentarily or permanently
inaccessible to the patient's conscience, and is therefore regarded as
\textit{unconscious\/} knowledge \cite{Fr66,Fr57}. In these considerations,
by \textit{symbolic representation\/} we mean the association 
of symbols to meaning, as occurs in language, as well 
as in other ways of expressing thoughts and emotions, such as artistic
representations (e.g., a painting or musical composition), 
and the recollection of dreams.

Consequently, an important part of 
the analytical treatment proposed by Freud consists 
of a procedure for ``{\it bringing what is unconscious 
into consciousness}''.  This is tantamount to
a basic technique aiming at ``\textit{filling up the gaps
in the patient's memories, to remove his amnesias\/}''~\cite{Fr66}.
Indeed, these amnesias are closely related to the origin of 
the neurotic  symptoms, that is, to the compulsion to repeat.
An important purpose of the therapeutic method called \textit{working-through\/},
whereby neurotic analysands obtain relief and cure 
of painful symptoms, is to develop knowledge regarding 
the causes of those symptoms. This is achieved by accessing 
unconscious memories through free associative talking, 
which yields understanding and  a change in
the analysand's compulsion to repeat~\cite{Fr57,Fr74,Fr53}.
The method is mainly based on the analysis of free associative talking,
symptoms, parapraxes (slips of the tongue and pen, misreading, forgetting, etc.), dreams,
and also on analyzing what is acted out in transference. This procedure allows the patient
to slowly symbolize his repressed memories and to create new representations
of his past experiences.

An illustrative, schematic model of this neurotic mechanism and the
\textit{working-through\/} therapeutic method, was advanced
in~\cite{We02,We09}. It was there proposed that the neuroses 
can be understood in terms of an associative memory process
in neural networks, where the network, when presented with an input 
pattern, retrieves a stored pattern which is most similar to
the one currently shown.
In order to model the compulsion to repeat a neurotic symptom, it was 
assumed that such a symptom is produced when the subject receives a
stimulus resembling a repressed or traumatic memory
trace. The stimulus then contributes to stabilize the neural net in a minimal energy
state, corresponding to the trace that synthesizes the original
repressed experience, which in turn produces a neurotic
response (an \textit{act\/}). The neurotic act does not result from
the stimulus as a new situation, but as a response to the repressed
memory.

The model is based on the conception that 
the linguistic, symbolic, and associative 
process associated with psychoanalytic 
working-through therapy is mapped
onto a corresponding scheme of reinforcing
synapses among memory traces in the brain.
These connections should involve declarative memory, 
implying that repressed memories are, at least partially,
transformed into conscious ones. This takes into
account the paramount importance that language has in psychoanalytic 
sessions, and the idea that unconscious memories are precisely 
those that cannot be expressed symbolically.
It was thus proposed that, as the analysand symbolically elaborates
manifestations of unconscious material through transference
in psychoanalytic sessions, he 
creates new neural connections, and reinforces or inhibits 
older ones, reconfiguring the topology of his
neural net.
The network topology arising from this reconfiguration process
stabilizes onto new energy minima, associated with new acts.

Following the memory organization
advanced in~\cite{We06} (see also~\cite{We09}),
it is assumed that neurons belong to two hierarchically
structured modules, respectively corresponding 
to {\em sensorial\/} and {\em symbolic memories\/}.
Mental images of stimuli received by 
sensory receptors, either from the environment or 
from the body itself, are represented by
memory traces stored in the sensorial memory module. 
On the other hand,  higher level representations
of traces in sensorial memory, {\em i.e. symbols\/},
are stored in the symbolic memory module. This
module represents brain structures corresponding to
symbolic processing, language, and consciousness.
Sensorial and symbolic memories are not isolated from 
each other and indeed, they interact, generating
unconscious and conscious mental activity.

In the model, the unconscious compulsion to repeat 
in neuroses~\cite{Fr66,Fr53,Fr74} is interpreted as a
bodily response (an act)  to an input stimulus (of any kind) 
that resonates with a pattern in sensorial memory, 
without activating symbolic memory. In this sense,
the compulsion to repeat is akin to a reflexive act. 
This accounts for neurotic patients' claim that
they cannot explain their neurotic acts.
A sensorial memory trace becomes conscious
when its retrieval also  activates the retrieval of
patterns in symbolic memory. If this happens, 
the resulting output does not resemble 
reflexive behaviour, and another level of processing 
becomes relevant. One may also have symbolic 
representations of emotions, as when someone says 
``I felt a warm happiness when I embraced my young nephew''.
Sensorial information that is not (or cannot be) associated
to a symbol stays unconscious. The neurotic mechanism
described here is in line with the hypothesis 
that the emergence of conscious experience
requires the existence of a physical (neural) 
layer that is able to support
metarepresentations~\cite{Cl07}.

A detailed description of the neural network model 
for these neurotic features can be found in~\cite{We09},
which includes a full description of the relevant algorithms.
Memory functioning was first modelled by recourse to a Boltzmann Machine (BM),
which is a stochastic extension of the Hopfield model.
However, it was later found that the node-degree distributions of the
hierarchically clustered network topologies generated by the 
model's \textit{clustering algorithm\/} (see section~\ref{sec:ModelNeuroses}), 
with long and short-range synapses, are well represented by
the asymptotic power-law, $q$-exponential distributions. This behaviour 
indicates that the statistical features of our model may not be well
described by Boltzmann-Gibbs statistical mechanics, but rather
by Nonextensive Statistical Mechanics (NSM)~\cite{Ts96,Ts09,Be09}.
The NSM theoretical framework, and its variegated applications to physics,
biology, economics, and other areas, have been the focus of an
intense research activity in recent years \cite{Ts09}. The concomitant
formalism has been applied to the study of diverse types
of complex systems, including systems with long-range interactions \cite{Vi11},
systems exhibiting weak chaos \cite{Ti16}, complex networks \cite{Br16},
processes involving nonlinear diffusion \cite{Ri11}, and many others.
In recent work~\cite{Ed13,Ed15} (see also references therein), 
the authors have also used Hopfield neural
networks to model attachment in developmental psychology, as well as behavioural patterns 
in psychology and psychotherapy.
In the present work, we thus use the model described in~\cite{We09}, where 
memory is simulated by a generalization of the BM
inspired on NSM, called Generalized Simulated Annealing (GSA)~\cite{Ts96,We09},
and this affects the sequence of associations of thoughts in the mental processes
we are illustrating. 

The model is in good qualitative agreement with the
main facts provided by psychoanalytic experience. 
In particular, it is consistent with the
(sometimes exasperating) slow nature of the working-through
process. The model's dynamics reproduces in a plausible way the re-association
of unconscious sensorial memory traces, and of new experiences, to symbolic processing areas,
mimicking the repetitive, adaptive, reinforcement learning involved in the simulation of working-through.
As a result of this self-reconfigura\-tion process,
represented in the model by a change in network connectivity,
the analysand  is partially freed from his/her original neurotic states 
and concomitant acts. The new network topology evolves to, and
stabilizes itself onto, new energy minima. These resulting network states
are associated with new conscious or unconscious
acts. Evidently, the ultimate aim of both the analysand and the analyst
is to reach new states, and generate new acts,
that are more pleasant and comfortable to the analysand and his relations.
As the therapy progresses, the analysand rewrites his/her 
life history, through a process of new significations. Furthermore,
and perhaps more importantly, both the present and the future of the analysand 
are also being rewritten, by creating new possibilities, and by opening 
new windows of opportunity for the pursuit of happiness.
Our central tenet is that this story is embedded, {\em i.e.\/} written,
in the individual's biological neural network.

It is clear that psychoanalytic theory is still
far from having the rigorous quantitative support 
required by modern science.
However, we agree with some contemporary scientists~\cite{Ka05,Sh10},
as well as early psychoanalysts~\cite{Fr66,La71,La91}
that, although this state of affairs poses serious limitations and difficulties, it also
constitutes a challenge and a stimulus for further scientific research.
These venues of enquiry are worth pursuing, since many findings
of psychodynamic theories have already contributed both to our understanding 
and characterization of mental phenomena, as well as to the development of successful 
clinical treatments for many mental disorders~\cite{Ka05,Sh10}.

In the present work, we study some additional properties of the aforementioned model, 
which give us further knowledge on how basic microscopic and macroscopic features and mechanisms 
influence emergent behaviour of the complex network structures proposed by the model. 
In Section~\ref{sec:ModelNeuroses}, we give a brief overview of the basic algorithms 
that characterize the model and which we use in our simulation experiments.
We present a definition of avalanche sizes during the memory retrieval mechanism and results of their 
measurements, in simulation experiments in Section~\ref{sec:Avalanches}. These avalanche sizes can 
be qualitatively compared to brain imaging experiments~\cite{Ta12}, showing that the NSM memory 
retrieval mechanism produces power-law behaviour, which does not emerge from
BM functioning, {\em i.e.\/} from Boltzmann-Gibbs statistical mechanics.

\section{\label{sec:ModelNeuroses} Network Topology and Memory Access Mechanisms}

The topological structure of each of the two memory modules was generated by a
\textit{clustering algorithm\/} proposed in~\cite{We06} (see also ~\cite{We09}), 
which models the self-organizing process 
that controls synaptic plasticity, resulting in a hierarchically structured 
neural network topology.
The algorithm is inspired on microscopic biological 
mechanisms, found in typical brain, cellular processes of many animals~\cite{Ha57,Ka00}.

As an example we can mention the \textit{on-center/off-surround\/} 
structure, characterized by neurons that are in cooperation, through 
excitatory synapses, with other neurons in their immediate neighbourhood, while 
they are in competition with neurons that lie outside these surroundings.
Competition and cooperation occur both in statically hardwired structures, 
and as part of a variety of neuronal dynamical processes, 
where neurons compete for certain chemicals~\cite{Ha57,Ka00}.  
In synaptogenesis, for instance, stimulated neurons release neural growth factors that 
spread through diffusion, reaching neighbouring cells and promoting synaptic 
growth. The cells receiving neural growth factors make synapses and live, 
while those having no contact with these substances die~\cite{Ka00,Ga03}. 
A neuron releasing neural growth factors ushers the process of synaptic 
formation in its tridimensional neighbourhood, and becomes a center of 
synaptic convergence. Neighbouring neurons releasing different 
neural growth factors at different rates give rise to several synaptic 
convergence centers. These centers then compete through the new synapses 
being created in their surroundings. Through these processes, a signaling network 
is established, that controls the development and plasticity of the 
brain's neuronal circuits. This neural competition is started and controlled 
by environmental stimulation and, consequently, it constitutes an important 
mechanism through which features of the environment can be mapped onto 
brain structures.

\subsection{Clustering Algorithm}

We reproduce the algorithm proposed in~\cite{We06,We09} here, to aid understanding of the measurements and
analysis which we introduce in this paper.
This \textit{clustering algorithm\/}  models the self-organizing process 
which controls synaptic plasticity, and results in a structured hierarchical
topology of each of the two memory modules. It consists of the following steps.
\begin{description}
  \item[Step 1] The initial bidimensional positions of the neurons are randomly 
                generated according to a uniform probability density on a square sheet.
  \item[Step 2] To simulate synaptic growth, we assume a Gaussian solution of the
                equation governing the diffusion of the neural growth factors, thus
                avoiding the time-consuming numerical treatment of this equation.
                Consequently, a synapse with strength $w_{ij}$ (the synaptic weight) 
                is allocated to transmit the output signal from neuron $n_j$ to a
                neuron $n_i$, according to the Gaussian probability density 
		\begin{equation}
		  P_{ij} = \exp (-(\mathbf{r_j} - \mathbf{r_i})^2 / (2\sigma^2) ) / \sqrt{2\pi\sigma^2}  \, ,
%		  P_{ij} = \frac{\exp (-(\vec{r_j} - \vec{r_i})^2 / (2\sigma^2) )}{\sqrt{2\pi\sigma^2}}  \, ,
                  \label{eq:gaussian}
		\end{equation}
                where $\mathbf{r_j}$ and $\mathbf{r_i}$ are the respective positions of $n_j$ and $n_i$ in the
                bidimensional sheet, and $\sigma$ is the standard deviation of the 
                Gaussian distribution, which is here a model
                parameter. If a synapse connecting $n_j$ to $n_i$ is generated, its
                strength $w_{ij}$ is proportional to $P_{ij}$.
  \item[Step 3] It was verified in~\cite{Ca03} that cortical maps representing different stimuli
                are formed, such that each stimulus activates a group of neurons
                spatially close to each other, and that these groups are uniformly
                distributed along the sheet of neurons representing memory.
                So one now randomly selects $m$ neurons which will each be a center of the representation
                of a stimulus. To choose the value of $m$, one should take into account the storage
                capacity of the BM~\cite{He91}.
  \item[Step 4] Reinforce synapses adjacent to each of the $m$ centers chosen in Step 3, 
                according to the following criteria. If $n_i$ is a center,
		define $sum_{n_i} =  \sum_j |w_{ij}|$.
                For each $n_j$ adjacent to $n_i$, increase $|w_{ij}|$ by $\Delta w_{ij}$,
                with probability $Prob_{n_j} = |w_{ij}| / sum_{n_i}$, where 
                $\Delta w_{ij} = \eta Prob_{n_j}$ and $\eta \in \Re$ is a model 
                parameter chosen in $[0, 1]$. After incrementing $|w_{ij}|$, decrement
                $\Delta w_{ij}$ from the weights of all the other neighbours of $n_i$,
                according to: $\forall k\neq j,  |w_{ik}| = |w_{ik}| - \Delta w_{ik}$,
                where  $\Delta w_{ik} = ( 1 - |w_{ik}| /  \sum_{k \neq j} |w_{ik}| ) \Delta w_{ij}$.
  \item[Step 5] Repeat step 4 until a clustering criterion is met. 

\end{description}

In the above clustering algorithm, Step 4 regulates the strength of synaptic connections, 
{\em i.e.\/}, \textit{plasticity\/}, 
by intensifying synapses within a cluster and reducing synaptic 
strength between clusters (disconnecting clusters). A cluster is therefore formed by 
a group of neurons that are close to each other, with higher probability of 
being connected by stronger synapses. This mechanism is akin to a preferential
attachment criterion, constrained by an energy conservation (neurosubstances) prescription, 
controlling synaptic plasticity. Neurons that have been sensorially more stimulated
and are therefore more strongly connected 
will stimulate other neurons in their neighbourhoods and promote still stronger connections. 
This is compatible with
the biological mechanisms mentioned earlier.

The growth of long-range synapses in the brain occurs less frequently than the
growth of short-range ones. The reason for this is that the former are energetically more costly 
than the latter. In order to allocate long-range synapses connecting clusters, one should 
regard the basic learning scheme proposed by Hebb~\cite{He91,Ka00a,Ed05}, which is based 
on the idea that synaptic growth among two neurons is promoted by their concomitant 
stimulation. Since we are still not
aware of the synaptic distributions that result in topologies which represent the structure 
of associations of symbols in language and thought,
as a first approximation, we have allocated long-range synapses randomly among clusters
of neurons (for a more detailed discussion see~\cite{We09,We06}).
Within a randomly chosen cluster $C$, defined by one of the $m$ neurons which is the center of 
representation of a stimulus (step 3 of the
clustering algorithm),  a neuron $n_i$ is chosen to receive a long-range connection
with probability $ P_i = \sum_j |w_{ij}| /  \sum_{n_j \in C} \sum_k |w_{jk}|$.
If the long-range synapse connects clusters in different memory sheets (sensorial
and symbolic memories), its randomly 
chosen weight is multiplied by a real number $\zeta$ in the interval $(0, 1]$, 
reflecting the fact that, in neurotic patterns, sensorial 
information is weakly accessible to consciousness, i.e., repressed.

\subsection{Memory Retrieval}

The topologies generated with the {\em clustering algorithm\/} 
have a hierarchical structure and the average node-degree distributions
present an asymptotic power-law behaviour~\cite{We09}. 
The functioning of memory retrieval was originally modelled by a Boltzmann Machine (BM)~\cite{He91,Ba93}.
There is no theoretical indication of the exact relation between network topology
and memory access dynamics. 
There have been indications that complex physical systems characterized by spatial 
disorder and/or long-range interactions, often presenting power-law behaviour 
(are asymptotically scale invariant) may be described by the 
Nonextensive Statistical Mechanics (NSM) formalism~\cite{Ts96,Ts09,Ri11,Vi11,Ti16,Br16,Be09}. 
The power-law and generalized $q$-exponential
behaviour for the node-degree distributions of the network 
topologies generated be the clustering algorithm~\cite{We09} indicate that they may not be well
described by Boltzmann-Gibbs (BG) statistical mechanics, but rather
by NSM~\cite{Ts96}.
Memory access was thus modelled by a generalization of the BM 
called Generalized Simulated Annealing (GSA)~\cite{We09,Ts96}, 
and this changes the chain of associations generated by the model.

In the BM~\cite{He91,Ba93}, the $N$ nodes in the neural network 
are connected symmetrically by weights
$w_{ij}=w_{ji}$. 
The state $S_i$ of each unit $n_i$ takes output values in $\{0, 1\}$.
As a consequence of the symmetry of the connections, one can associate an energy functional
\begin{equation}
  H(\{S_i\})=-\frac{1}{2}\sum_{ij}w_{ij}S_i S_j \; ,
  \label{eq:BGEnergy}
\end{equation}
to network state $S = \{S_i\}$ and, 
according to the BG distribution, the 
transition probability (acceptance probability) from state $S$ to $S'$,
if $H(S') \geq H(S)$, is given by
\begin{equation}
  P_{BG}(S \rightarrow S')=\exp \left[\frac{H(S)-H(S')}{T}\right] \; ,
\label{eq:BGTransProb}
\end{equation}
where $T$ is the network ``temperature'' parameter.
Pattern retrieval on the net is achieved by a standard simulated annealing 
process, in which $T$ is gradually lowered by a factor
$\alpha$, according to the BG distribution~\cite{He91}. 
In the NSM formalism, one uses a generalized acceptance probability~\cite{Ts96} 
for a transition from $S$ to $S'$, if $H(S') \geq H(S)$, given by
\begin{equation}
  P_{GSA}(S \rightarrow S') = 
  \begin{cases}
     \frac{1} {\left[1 + (q_A - 1) (H(S') - H(S)) / T \right]^{1/(q_A-1)} }\, , & \text{\scriptsize {if $\varphi  > 0$ }}\, , \\
     0\, , & \text{\scriptsize {if $\varphi \le 0$}}\; ,
%     \frac{1} {\left[1 + (q_A - 1) (H(S') - H(S)) / T \right]^{1/(q_A-1)} }\, , & {\tiny \text{if  }}{\scriptscriptstyle 1 + (q_A - 1) (H(S') - H(S)) / T  > 0}\, , \\
%     0\, , & {\tiny \text{if }} {\scriptscriptstyle 1 + (q_A - 1) (H(S') - H(S)) / T \le 0}\; ,
  \end{cases}
\label{eq:TsaTransProb}
\end{equation}
where $q_A$ is a parameter called $q$-acceptance and $\varphi = 1 + (q_A - 1) (H(S') - H(S)) / T$. 
In the limit $q_A \to 1$, (\ref{eq:TsaTransProb}) reduces to the Boltzmann-Gibbs transition probability (\ref{eq:BGTransProb}).
If one uses transition probability~(\ref{eq:TsaTransProb}) in place of transition 
probability~(\ref{eq:BGTransProb}), in the standard BM simulated annealing algorithm, the resulting
procedure is called GSA (see~\cite{Ts96} for a more detailed discussion). 

It is convenient that we define, for each transition from state $S$ to $S'$ during annealing, the quantity
\begin{equation}
  \Delta E = H(S') - H(S) \; .
\label{eq:DeltaE}
\end{equation}

Both the BM and GSA differ from a gradient descent minimization scheme since,
besides allowing state transitions that lower the total energy of the 
network~(\ref{eq:BGEnergy}), they also allow the system to transition into a state 
with an increase in energy, depending on the
values of $T$ and $q_A$, according to~(\ref{eq:BGTransProb}) and~(\ref{eq:TsaTransProb}). 
The BG transition probability~(\ref{eq:BGTransProb}) predominantly allows changes of states with small 
increases in energy, and state transitions with higher energy increases occur with almost 
negligible probabilities. 
The BM will thus strongly prefer visiting state space within a nearby energy neighbourhood 
from the starting point (initial state of the annealing process). The GSA transition 
probability~(\ref{eq:TsaTransProb}) allows state transitions with higher
energy increases than the BM and, although the probabilities for these 
transitions are very low, they are still considerable when compared to the BM. This allows the system to
transition into attractor states that are farther in state space from the initial
network state and also into basins of attraction corresponding to higher energy values~\cite{We09,We11}
(see Figure 2 in~\cite{We11}).
This implies that the hierarchically structured memory modules having both long and short-range 
connections, with memory access modelled by GSA achieves associations among memory traces 
that are not achieved by the BM~\cite{We11}. This increase in associativity observed in the GSA 
memory retrieval mechanism, when compared to the BM, suggests that if memory functions according to
the NSM theoretical framework, one will have a more creative mode of memory and mental functioning. 

In traditional neural network modelling, the temperature parameter is inspired by the fact that 
biological neurons fire with variable strength, and that there are delays in synapses, random
fluctuations from the release of neurotransmitters, and so on. 
These effects can be considered as noise in synaptic functioning~\cite{He91,Ka00},
and we may thus consider that temperature in the BM and GSA and the $q_A$ parameter 
in GSA control noise. 
In the model we are considering~\cite{We09}, non-zero temperature and $q_A \ne 1$ values
regulate associations among memory traces, in an analogy with the concept that freely talking 
in analytic sessions and stimulation from the psychoanalyst lower resistances and 
allow greater associativity and creativity.

Once the network topology is generated by the \textit{clustering algorithm\/}, 
one can find the stored patterns by presenting
many random patterns to both the BM and the GSA mechanism, with an annealing schedule 
$\alpha$ that allows stabilizing onto the many local minimum values of the
network energy function. Each of the minimum energy values corresponds to
a stable state of the network and is associated with a stored memory trace.
These initially stored patterns represent the neurotic memory attractors 
(as in the compulsion to repeat), since they are associated with the two weakly 
linked sensorial and symbolic subnetworks.
In~\cite{Ca03}, Carvalho \textit{et al.\/} proposed a neurocomputational model to describe how 
the original memory traces are formed in cortical maps.
It is important to note that in our experiments, we are not aiming at finding a global 
minimum energy state, but at visiting the many local minima of the energy landscape, 
which represent stored information in a cognitive network.

\section{\label{sec:Avalanches} Avalanches in Memory Access}

In the present work, we have defined and measured avalanche sizes during the memory retrieval 
process. The size of an avalanche during the simulated annealing 
procedure, both in the BM and in GSA, is defined to be the number of state changes 
that occur during annealing, from the initial to the final
network state, which corresponds to one of the minima in the energy
landscape of the network. This is equivalent to saying that the size of an avalanche corresponds to the 
number of hops (steps) that the access mechanism performs on the network energy landscape, 
from an initial state until it reaches one of the local minima. 
The avalanche size should thus be proportional to the time for the propagation of a signal (stimulus),
during access of information in memory.
In other words, the avalanche size should be related to the time
associated with memory retrieval and also to the size of the neuronal 
region activated during memory access.

In each of the simulation experiments we present here, we performed 2,048,000 
minimization (annealing) procedures, 
starting each one from a different random network configuration. Since the simulations 
require much computational time (many days and even a few weeks), 
we have analysed smaller networks with total number 
of neurons $N = 32$, such that $N_{sens} = N_{symb} = 16$ neurons belong to the sensorial and symbolic modules,
respectively. This is a small network size when one considers the brain, as there are 
billions of neurons in the human brain. However,
the brain is considered to be scale invariant~\cite{Ta12} and because the short-range mechanisms
in the algorithms we use to create network topology are scalable, we expect that our experiments 
should, to some extent, qualitatively be comparable to biological processes~\cite{We09}.

The annealing schedule was controlled so that the network stabilizes on
the many local minima of the energy landscape, and was the same for both machines in all experiments. 
We conducted experiments with different values of the temperature $T$ and $q_A$ parameters. 
Both of these parameters model stochastic fluctuations in network functioning and we studied
how the change in $T$ and $q_A$ values caused a change in the behaviour of the two machines, thus
affecting the distribution of avalanche sizes, the network energy loss and measurements of correlations
among energy increments along the path followed during annealing.

\begin{figure}[ht]
\centering
\begin{tabular}{cc}
  {\includegraphics[height=5.6 cm]{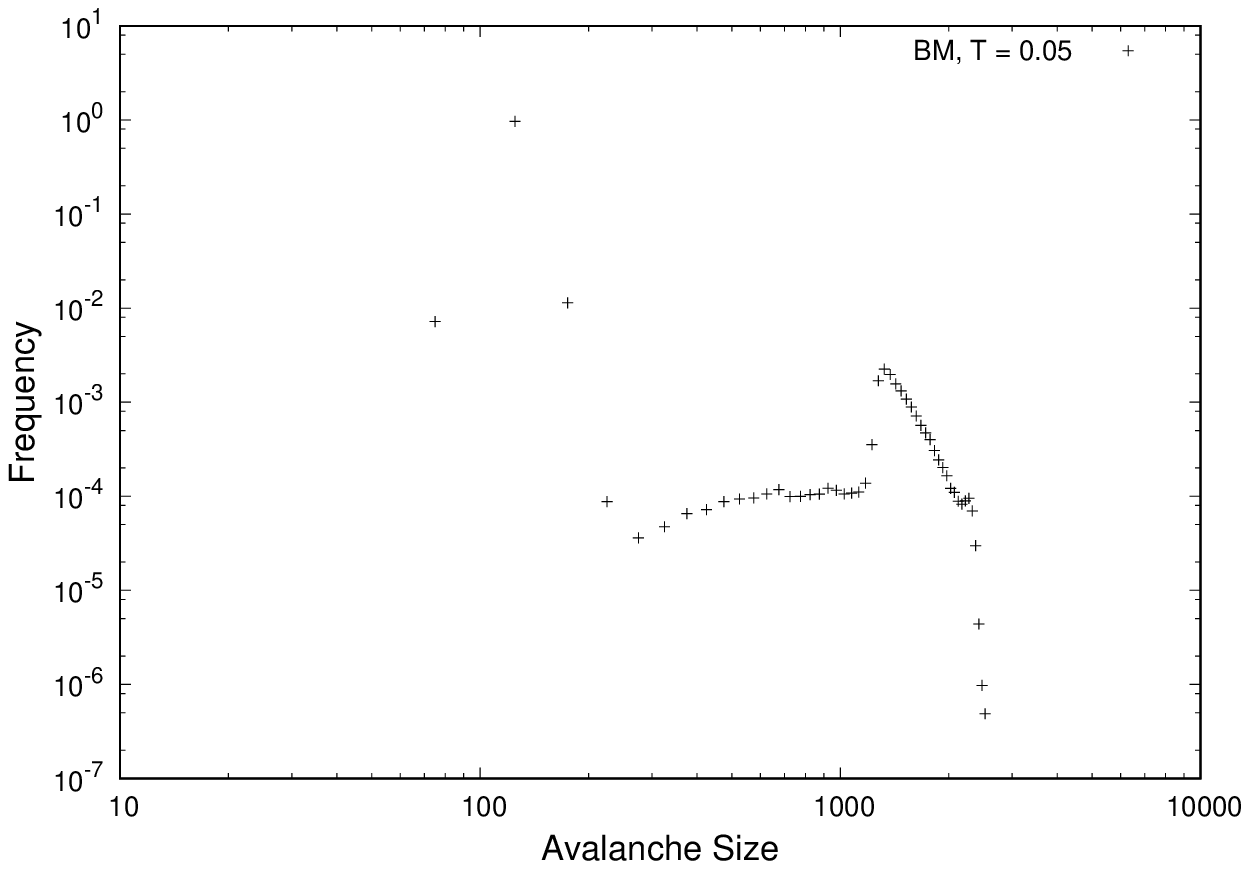} }
&
  {\includegraphics[height=5.6 cm]{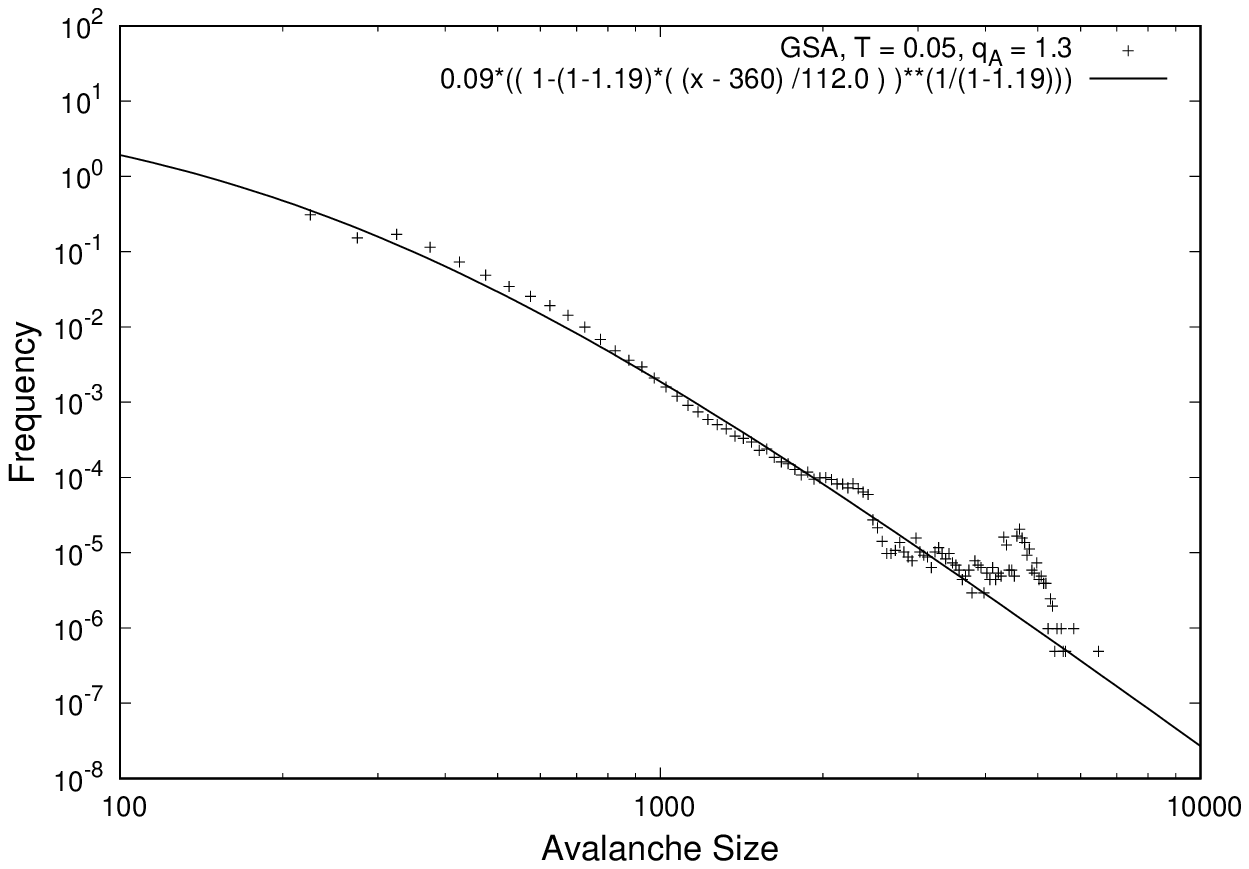} }\\
  (\textbf{a})&(\textbf{b})
 \end{tabular}
%   \vspace{-24pt}
  \caption{ (\textbf{a}) Frequency of occurrence of avalanche sizes for the BM.
  (\textbf{b}) Frequency of avalanche sizes with GSA. For
    both machines, $T = 0.05$ and for GSA, $q_A = 1.3$. 
    In both cases, 50\% of the synapses are inhibitive. All depicted quantities are
    dimensionless.}
  \vspace{0.8cm}
  \label{Fig:AvalSizes_50Inhib_T_0_05_q_1_3}
\end{figure}

In Figure~\ref{Fig:AvalSizes_50Inhib_T_0_05_q_1_3}, we show the avalanche size distributions obtained with
the two machines, for $T = 0.05$ and $q_A = 1.3$. In this experiment, when executing 
step 2 of the clustering algorithm, 50\% of the synapses
of the network are excitatory (positive synaptic weights) and 50\% are inhibitory
(negative synaptic weights). On the left, we see the avalanche size distributions for
the BM and on the right, for GSA. From the distribution in Figure~\ref{Fig:AvalSizes_50Inhib_T_0_05_q_1_3}-a, 
it is clear that the BM has a preferred avalanche size of 125 with an associated probability of occurrence of 0.96.
This distribution also has a smaller peak at avalanche size 1325, which occurs with probability 0.0022.
The smaller peak has a slower decay before a sharp drop.

The distribution for GSA in Figure~\ref{Fig:AvalSizes_50Inhib_T_0_05_q_1_3}-b, on the other hand, 
behaves quite differently. For the range of avalanche sizes we measured, we observe a monotonic 
drop in the frequency for GSA. The points measured in this simulation can be approximately 
fitted by a $q$-exponential~\cite{Ts09}
\begin{equation}
  \exp_q(x) =
  \begin{cases}
     [1 + (1-q) x]^{\frac{1}{1-q}}\, , & \text{if  } 1 + (1-q)x > 0\, , \\
     0\, , & \text{if }  1 + (1-q)x \le 0\, .
  \end{cases}
  \label{eq:qexp}
\end{equation}
This function is at the core of nonextensive thermostatistics, being the result
of the constrained optimization of the power-law, nonadditive entropic functional 
\begin{equation}
   S_q = \frac{1}{q - 1}[1 - \sum_i p_i^q]\, ,
   \label{eq:Sq}
\end{equation}
where $p_i$ is a normalized probability distribution and $q$ is a real parameter. 
The $q$-exponential asymptotically behaves as a power-law and, in 
Figure~\ref{Fig:AvalSizes_50Inhib_T_0_05_q_1_3}-b, the curve which fits the points within
the observed values of avalanche sizes,
corresponding to $q = q_f = 1.19$, does show an asymptotic behaviour akin to a power-law.

%\begin{figure*}[ht]
%\centerline {
%  {\includegraphics[height=6.5 cm]{Aval_Size_BM_LogFreq_LogAvSize_50_Inhib_q_1_3_T_0_05} }
%   \hfil
%  {\includegraphics[height=6.5 cm]{Aval_Size_GSA_LogFreq_LogAvSize_50_Inhib_q_1_3_T_0_05} } }
%   \caption{(a) On the left, frequency of avalanche sizes for the Boltzmann Machine. 
%   (b) On the right, frequency of avalanche sizes with Generalized Simulated Annealing. 
%   In both cases, 50\% of the synapses are inhibitive.}
%   \label{Fig:AvalSizes_50Inhib}
% \end{figure*}
%%%%%%%%%%%%%%%%%%%%%%%%%%%%
%%%%%%%%%%%%%%%%%%%%%%%%%%%%

The imaging technique in neuroscience referred to as functional magnetic resonance imaging (fMRI) uses
a method known as {\em blood-oxygen-level dependent\/} (BOLD) contrast imaging~\cite{Hu09}. Most cells in the body
possess their own reservoirs of sugar which undergoes respiration to produce energy, when the cells need it.
It is known that neurons do not possess internal stores of energy and, when a neuron is in a firing state it requires
energy in the form of sugar and oxygen, which it obtains by means of a haemodynamic response, whereby
blood releases oxygen to the firing neurons at a greater rate in comparison to those in resting state. This
causes a difference in magnetic susceptibility between oxygenated and deoxygenated blood, which results in a
magnetic signal variation that may be detected on an MRI scanner~\cite{Hu09}. Studying changes in the brain
BOLD signal allows observation of different active areas of the brain to further understand 
certain aspects of brain functioning.

In 2003, the work of Beggs and Plenz~\cite{Be03} with cortical slices 
revealed the phenomena which they called neuronal
avalanches in the brain. In those observations, avalanches were defined as short bursts of activity 
that last a few milliseconds, followed by several seconds of inactivity. 
These phenomena can be observed in cortical slices of the neocortex, although
their relation to physiological processes in the brain is still unknown~\cite{Be03}. 
The avalanche sizes detected by Beggs and Plenz follow a power-law-like distribution, with an exponent of -3/2. 
In their observations, they recorded spontaneous local field potentials continuously using a 60 channel multielectrode array, 
and avalanches are a measure of time duration and
spatial reach of the propagation of signals in the brain.

Tagliazucchi {\em et al.\/}~\cite{Ta12} extended the work done by Beggs and Plenz~\cite{Be03}, 
``by inspecting only the relatively large amplitude BOLD signal peaks, suggesting
that relevant information can be condensed in {\em discrete events\/}''.
The authors present a {\em spatiotemporal point process\/}, where the timing and location of these
discrete events is obtained to study and capture the dynamics of the brain in a resting state. When regions of the brain
show activity above a certain threshold level, the detected BOLD signal determines the location of the activity as well
as the duration. Figure 3-A in~\cite{Ta12} shows examples of co-activated {\em clusters\/},  
defined as ``... groups of contiguous voxels with signal above the threshold at a given time'' 
(clusters are measured in units of voxels).

In~\cite{Ta12}, Figure 3-D shows the distribution of average measured cluster sizes 
in fMRI images for ten individuals, 
which has a power-law (scale-free) behaviour spanning four orders of magnitude, also with an exponent of approximately -3/2. 
The size of a cluster is proportional to the duration of brain activity. The
longer the duration for which an avalanche is seen in the brain, the larger are the observed cluster sizes. 
This happens because when the duration of activity is longer, clusters (voxels) 
cause the activation of neighbouring clusters, before reducing activity and fading away. 
This behaviour is comparable to the avalanche size distributions produced by our computational model,
as in Figure~\ref{Fig:AvalSizes_50Inhib_T_0_05_q_1_3}-b,
because an avalanche size, in our simulation experiments, was defined in a way so that it should be proportional 
to the time for propagation of a signal during the access of information in
memory. This is a reasonable assumption, since transition probabilities~(\ref{eq:BGTransProb}) and~(\ref{eq:TsaTransProb})
which regulate network functioning should mimic the way the memory network occupies
phase-space, during the retrieval of a mnemenic trace.
So the larger avalanche sizes observed in our simulations should be associated with larger cluster sizes, 
as a larger avalanche size is associated with a longer time for propagation of a signal, with access to a 
larger number of neighbouring neurons. We therefore {\em qualitatively} compared the behaviour of the 
distribution of cluster sizes obtained in~\cite{Ta12} with the distribution of avalanche sizes 
obtained in our simulation experiments, for some values of the $T$ and $q_A$ parameters,
and we found $q$-exponentials with a similar asymptotic power-law like behaviour, spanning similar 
orders of magnitude of avalanche sizes and frequencies, only for the GSA machine, however with 
larger absolute values of the exponent (exponent close to -5 in Figure~\ref{Fig:AvalSizes_50Inhib_T_0_05_q_1_3}-b and
approximately -10 in Figure~\ref{Fig:AvalSizes_30Inhib_T_0_2_q_0_7}-b). The Boltzmann machine 
did not present power-laws in our experiments.
This qualitatively corroborates our initial indication, that the power-law and generalized $q$-exponential
behaviour of quantities which describe the structure of the complex networks generated by 
the clustering algorithm of our model suggest that they may not be well
described by Boltzmann-Gibbs (BG) statistical mechanics, but rather
by NSM~\cite{Ts96,We09}.

For a sequence of values of energy increments $\Delta E$ given by~(\ref{eq:DeltaE}) during an avalanche, 
we also measured if the value $\Delta E(\tau_0)$ at step $\tau_0$ influences (is correlated to) the value 
at a later step $\Delta E(\tau_0 + \tau)$.
The temporal correlation function $G(\tau)$ among elements in a sequence separated by an interval $\tau$ 
is defined as~\cite{Je98} 
\begin{equation}
   G(\tau) = \sum_{\tau_0 = 1}^{m - \tau} \frac{\Delta E (\tau_0) \Delta E (\tau_0 + \tau)}{m - \tau} -
             \left( \sum_{\tau_0 = 1}^{m} \frac{\Delta E (\tau_0)}{m} \right)^2    \, ,
   \label{eq:Correlation}
\end{equation}
where $m$ is the size of the sequence which is being considered, and should be much larger than the largest
value of $\tau$ in the domain of~(\ref{eq:Correlation}).

\begin{figure}[ht]
\centering
\begin{tabular}{cc}
  {\includegraphics[height=5.6 cm]{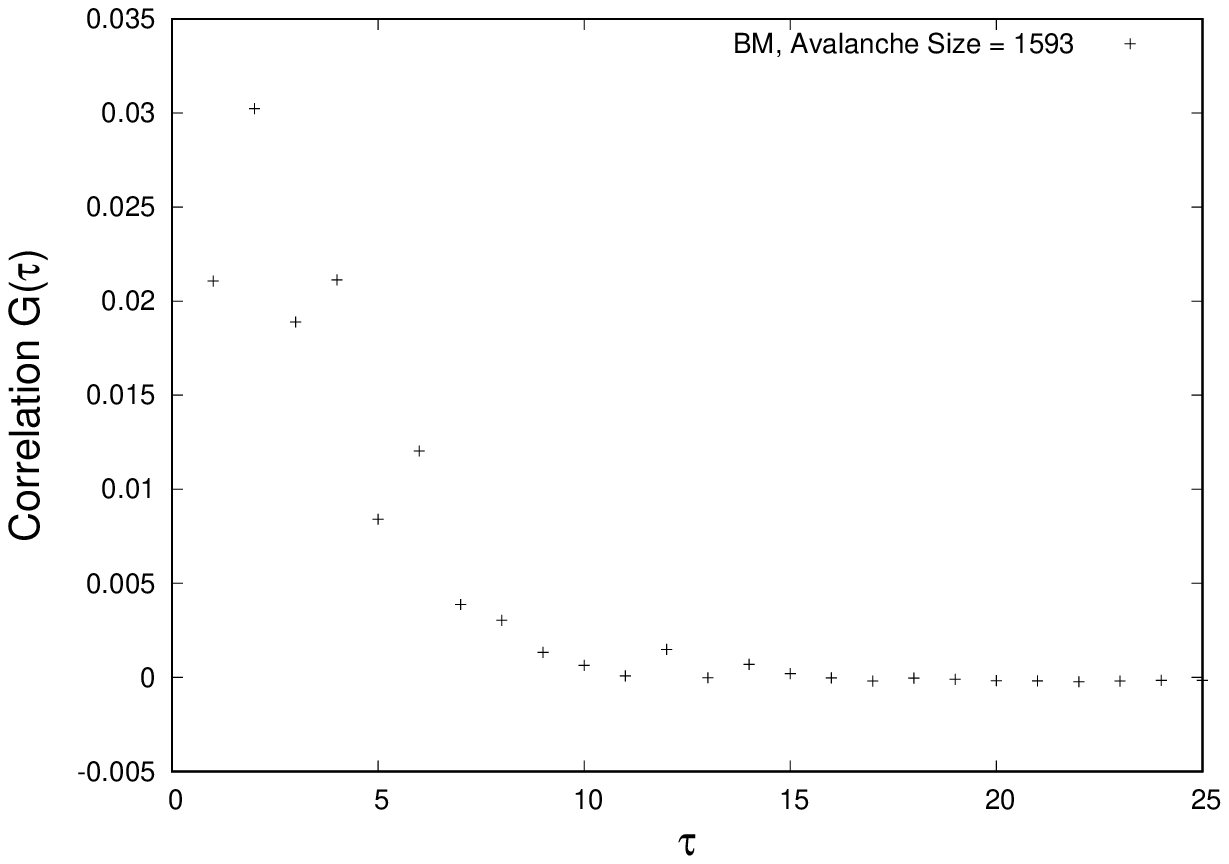} }
&
  {\includegraphics[height=5.6 cm]{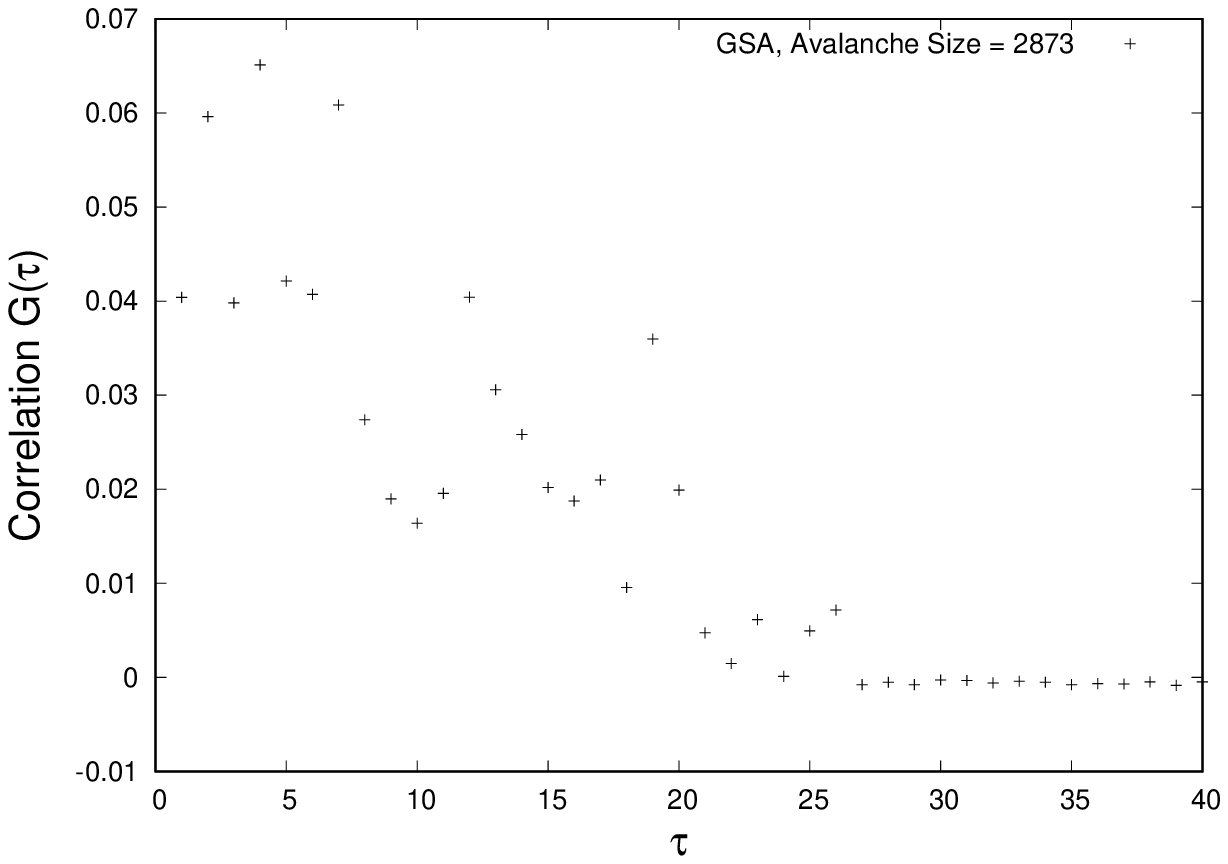} }\\
  (\textbf{a})&(\textbf{b})
 \end{tabular}
%   \vspace{-24pt}
  \caption{ (\textbf{a}) Correlations among the sequence of $\Delta E$ values during an avalanche of the BM. 
              The avalanche size for this experiment is $m = 1593$ and the maximum value of $\tau$ is 25. 
  (\textbf{b}) Correlations among the $\Delta E$ for an avalanche generated by GSA. The avalanche
              size for this experiment is $m = 2873$ and the maximum value of $\tau$ is 40. For
              both machines, $T = 0.05$ and for GSA, $q_A = 1.3$. 
              50\% of the synapses are inhibitive. All depicted quantities are
              dimensionless.}
  \vspace{0.8cm}
  \label{Fig:Correlations_50Inhib_T_0_05_q_1_3}
\end{figure}

In Figure~\ref{Fig:Correlations_50Inhib_T_0_05_q_1_3}, we show correlations
for one avalanche of the BM in Figure~\ref{Fig:Correlations_50Inhib_T_0_05_q_1_3}-a 
and for another avalanche generated by GSA in Figure~\ref{Fig:Correlations_50Inhib_T_0_05_q_1_3}-b.
Since avalanches in the BM and GSA typically have different sizes, 
Figures~\ref{Fig:Correlations_50Inhib_T_0_05_q_1_3}-a and b correspond to
different values of $m$. For both machines we notice an approximate exponential fall of $G(\tau)$ for
smaller values of $\tau$, with larger fluctuations around an average decay for GSA. These larger fluctuations for GSA
occur because GSA generates higher probabilities of making state transitions with positive 
values of $\Delta E$ than the BM. After a rapid decrease,
both figures converge to a situation where they fluctuate around zero correlation values. 
A similar behaviour occurs for the other avalanches which we measured. For larger values of
$T$ and $q_A$, the fluctuations around average correlation values are much larger, since the machines 
then have higher probabilities of making state transitions with positive values of $\Delta E$,
and the first term in~(\ref{eq:Correlation}) is negative more often than for lower values of $T$ and $q_A$.
%%%%%%%%%%%%%%%%%%%%%%%%%%%%
%%%%%%%%%%%%%%%%%%%%%%%%%%%%

\begin{figure}[ht]
\centering
\begin{tabular}{cc}
  {\includegraphics[height=5.6 cm]{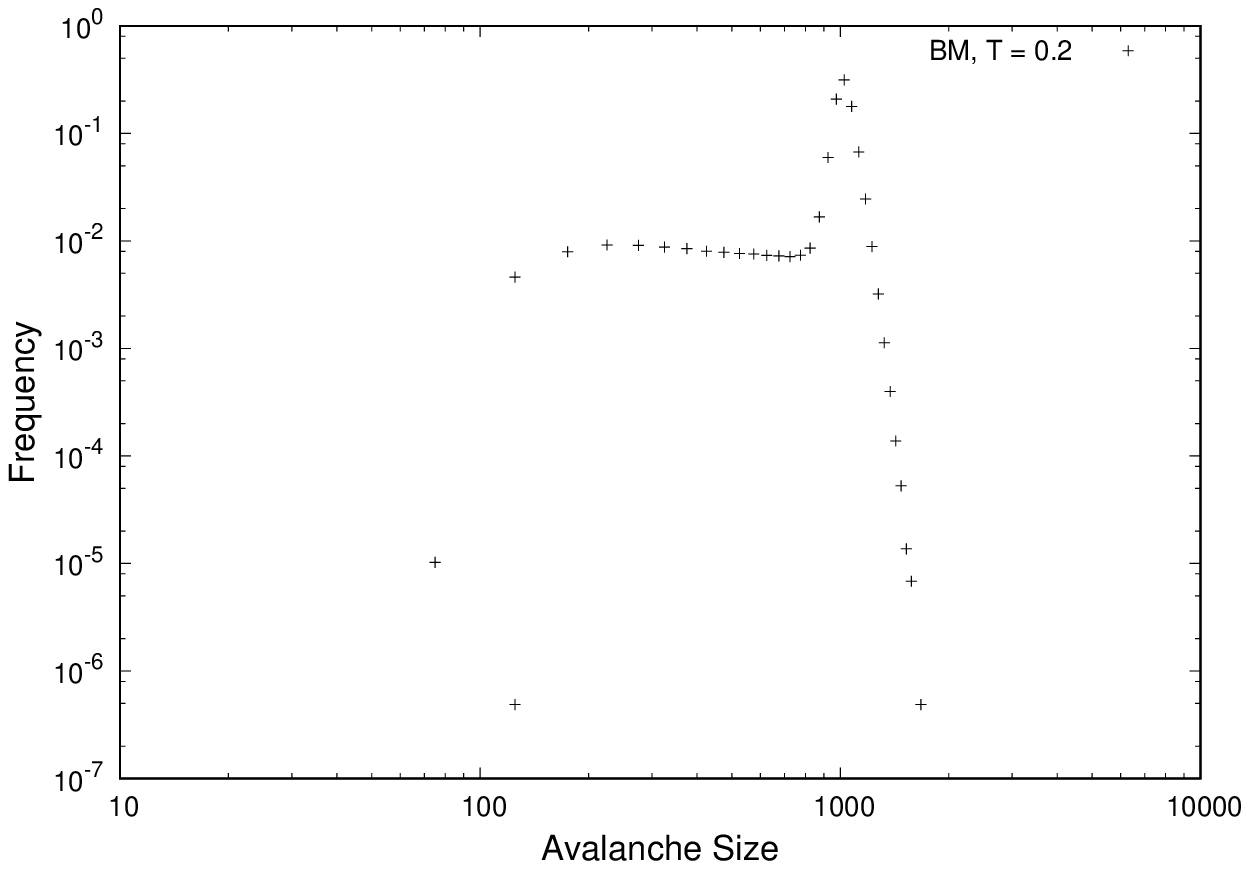} }
&
  {\includegraphics[height=5.6 cm]{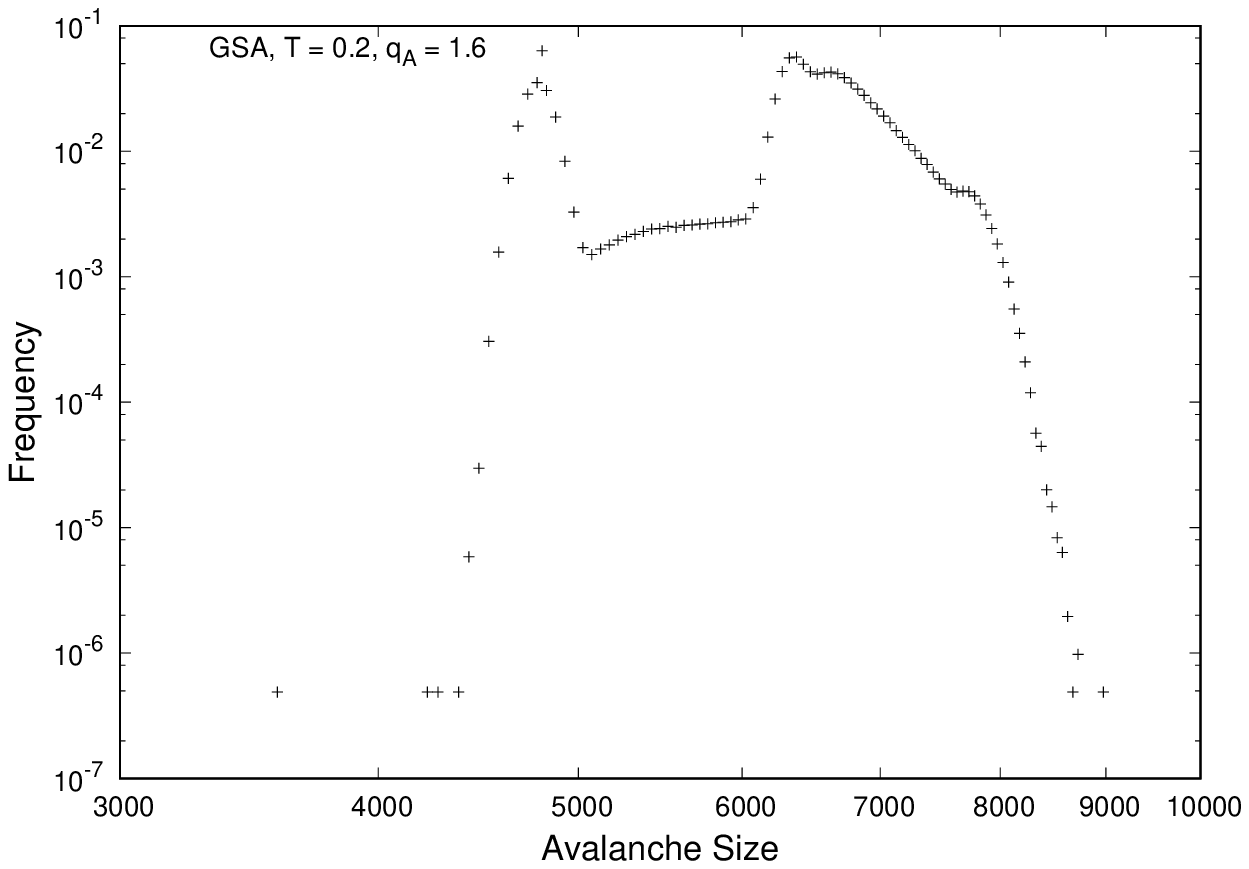} }\\
  (\textbf{a})&(\textbf{b})
 \end{tabular}
%   \vspace{-24pt}
  \caption{ (\textbf{a}) Frequency of occurrence of avalanche sizes for the BM.
  (\textbf{b}) Frequency of avalanche sizes with GSA. For
    both machines, $T = 0.2$ and for GSA, $q_A = 1.6$. 
    In both cases, 50\% of the synapses are inhibitive. All depicted quantities are
    dimensionless.}
  \vspace{0.8cm}
  \label{Fig:AvalSizes_50Inhib_T_0_2_q_1_6}
\end{figure}

\begin{figure}[ht]
\centering
\begin{tabular}{cc}
  {\includegraphics[height=5.6 cm]{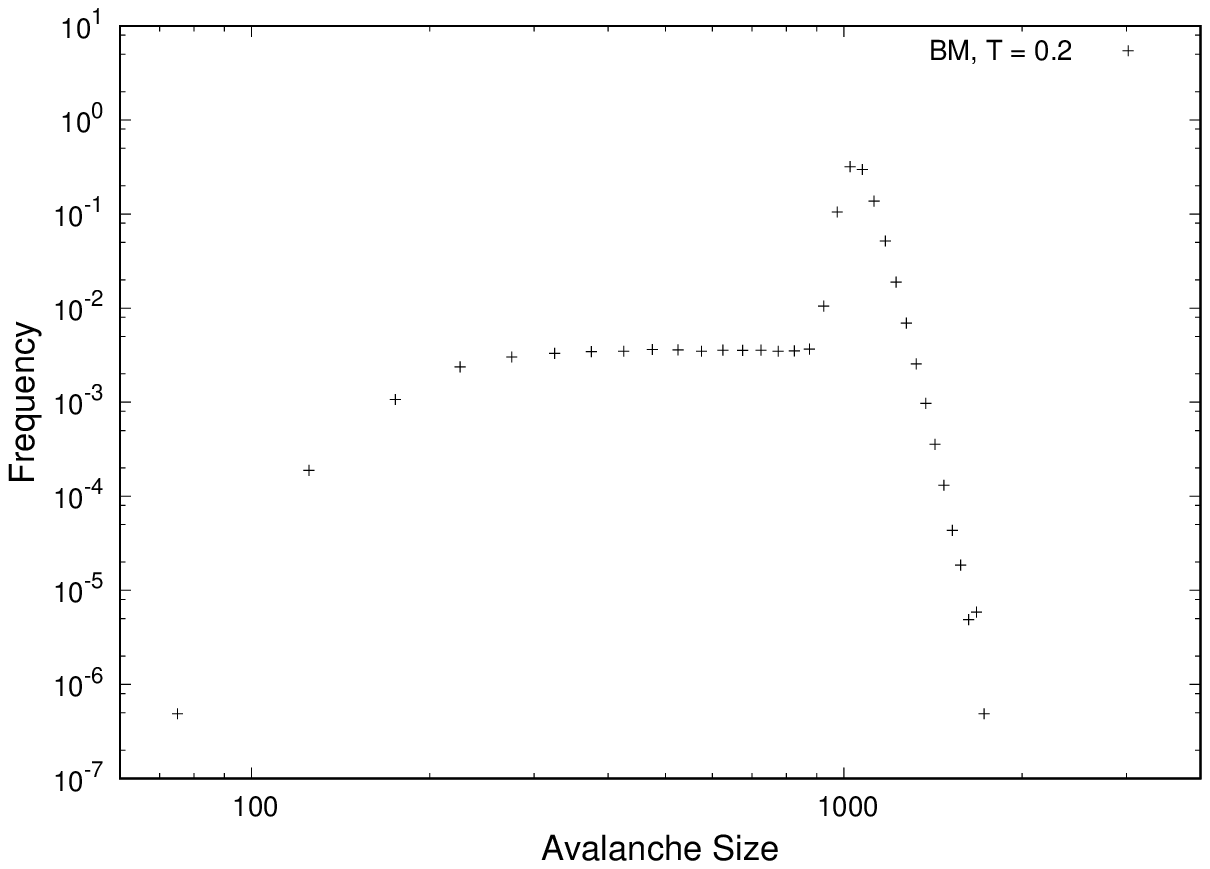} }
&
  {\includegraphics[height=5.6 cm]{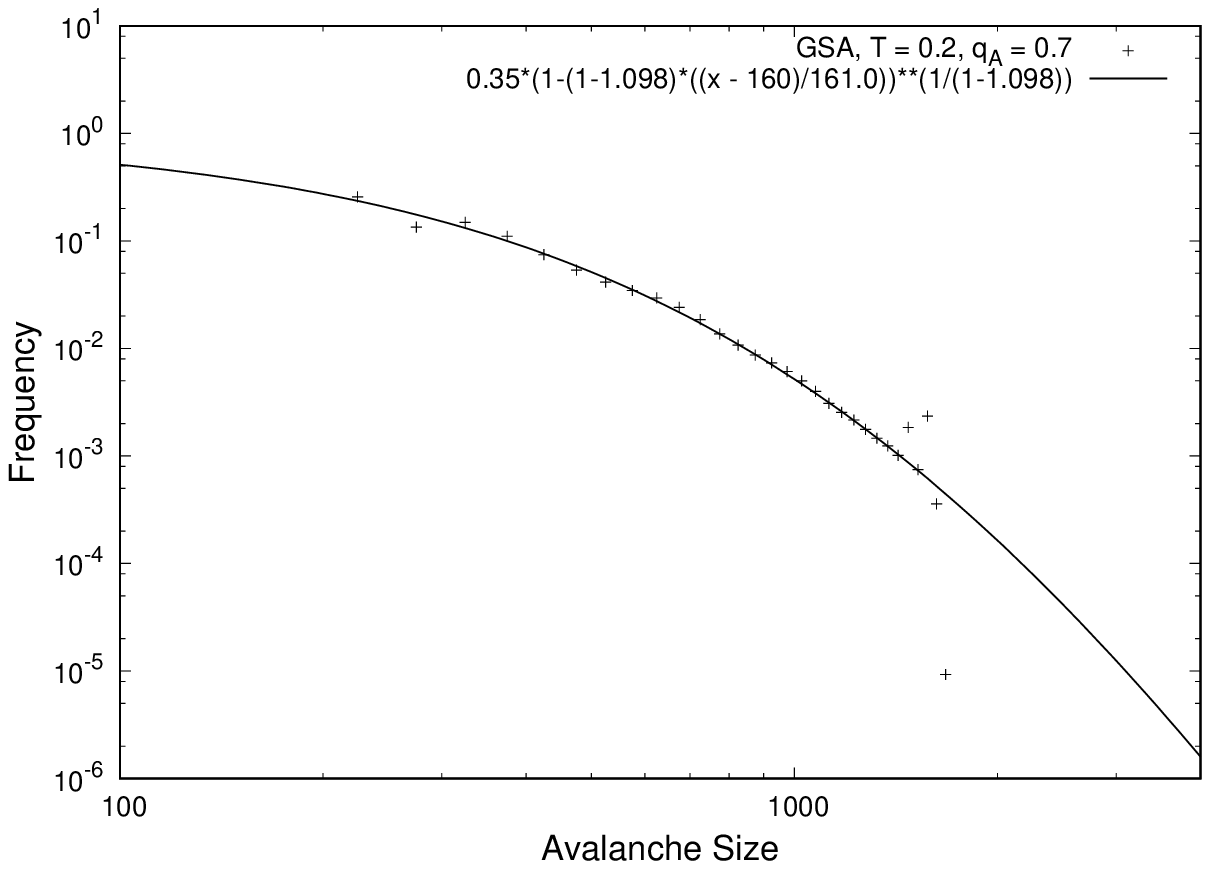} }\\
  (\textbf{a})&(\textbf{b})
 \end{tabular}
%   \vspace{-24pt}
  \caption{ (\textbf{a}) Frequency of occurrence of avalanche sizes for the BM.
  (\textbf{b}) Frequency of avalanche sizes with GSA. For
    both machines, $T = 0.2$ and for GSA, $q_A = 0.7$. 
    In both cases, 30\% of the synapses are inhibitive. All depicted quantities are
    dimensionless.}
  \vspace{0.8cm}
  \label{Fig:AvalSizes_30Inhib_T_0_2_q_0_7}
\end{figure}

We show avalanche size distributions obtained with
the two machines, for $T = 0.2$ and $q_A = 1.6$, in Figure~\ref{Fig:AvalSizes_50Inhib_T_0_2_q_1_6}.
As a consequence of the increase of the values of the two parameters that control associativity
and noise in the network, both machines have a higher probability of increasing energy during the
annealing process, and therefore achieve higher values of avalanche sizes more frequently.
The BM now has a preferred avalanche size of 1025, with a corresponding probability of 0.31, as seen 
from Figure~\ref{Fig:AvalSizes_50Inhib_T_0_2_q_1_6}-a. The distribution of avalanche sizes 
for GSA now loses the general $q$-exponential behaviour and has two preferred values 
(the first is 4775 with 0.035 probability and the second is 6375 with probability 0.056),
with a very short power-law like behaviour after the
second peak, before a sharp drop. The increase in frequency of larger avalanche sizes for
both machines is consistent with the increase in associativity for larger values of $T$ and $q_A$ found 
in~\cite{We11}, with GSA still producing more associativity than the BM.

%%%%%%%%%%%%%%%%%%%%%%%%%%%%
%%%%%%%%%%%%%%%%%%%%%%%%%%%%

There has been recent work regarding the study of the proportion of inhibitory synapses in
the brain (see~\cite{Ca15} and references therein). In mammals, this proportion has been measured
to have values between 20 and 30\%. We have thus conducted avalanche measurements in our model
so that, when executing  step 2 of the clustering algorithm, 70\% of randomly chosen synapses
of the network are excitatory (positive synaptic weights) and 30\% are inhibitory
(negative synaptic weights). A distribution of avalanche sizes for this proportion of inhibitory
synapses, for both the BM and GSA with $T = 0.2$, and $q_A = 0.7$ in GSA is shown
in Figure~\ref{Fig:AvalSizes_30Inhib_T_0_2_q_0_7}. Again we see a preferred avalanche size for
the BM, in this case at 1025 with probability 0.32. We also observe once more for GSA in 
Figure~\ref{Fig:AvalSizes_30Inhib_T_0_2_q_0_7}-b, for the range of avalanche sizes which we measured, a monotonic 
decrease in the distribution of avalanche sizes, so that the experimental points resulting from this simulation 
can be approximately fitted by a $q$-exponential, with $q = q_f = 1.098$. This is also asymptotically  
in accordance with the scale-free, power-law behaviour observed experimentally in~\cite{Ta12}.

%%%%%%%%%%%%%%%%%%%%%%%%%%%%
%%%%%%%%%%%%%%%%%%%%%%%%%%%%

%\begin{figure*}[ht]
%\centerline {
%  {\includegraphics[height=6.5 cm]{Aval_Size_BM_LogFreq_LogAvSize_30_Inhib_q_0_7_T_0_2} }
%   \hfil
%  {\includegraphics[height=6.5 cm]{Aval_Size_GSA_LogFreq_LogAvSize_30_Inhib_q_0_7_T_0_2} } }
%   \caption{(a) On the left, frequency of avalanche sizes for the Boltzmann Machine. 
%    (b) On the right, frequency of avalanche sizes with Generalized Simulated Annealing. 
%    In both cases, 30\% of the synapses are inhibitive.}
%   \label{Fig:AvalSizes_30Inhib}
%\end{figure*}

Besides the avalanche size distributions and temporal correlations, we also measured 
the distributions of the total energy lost by the network, during the annealing 
processes of each set of simulation experiments, for both machines, {\em i.e.\/}
$H(S_{Final}) - H(S_{Init})$ , where $S_{Final}$ and $S_{Init}$ are respectively the final and initial
states of the network, during one memory retrieval procedure. These distributions are basically the same for
both machines during the same experiment, with just a very slight and negligible difference close to the tales
of the distributions, due to limited size statistics.

\section{\label{sec:Concls} Conclusions}

We have further studied the memory mechanism in a schematic neural network model, 
which illustrates conscious and unconscious mental processes in neurotic 
mental functioning~\cite{We09,Fr74,Fr53,Fr66,Fr57}. The model illustrates
aspects of mental functioning related to memory associativity, creativity,
consciousness and unconsciousness, which are present both in pathological 
and normal mental processing.
The model is based on a self-organizing, clustering algorithm which generates a 
hierarchically structured bimodular neural network, based on basic
biological cellular mechanisms, which have aspects akin to a preferential attachment
mechanism, constrained by an energy conservation (conservation of neurosubstances) 
prescription, controlling synaptic plasticity. 
The networks generated by the algorithm have a small-world-like 
topology, with clusters of neurons strongly coupled by short-range connections (synapses) and 
also less frequent long-range connections. The node-degree distributions of these
network topologies present a generalized $q$-exponential and asymptotic power-law behaviour.
This structure suggests that memory access may best be modelled by a generalization of the
Boltzmann Machine called Generalized Simulated Annealing, and this determines the chain of
associations generated by the model.

In the present work, we have introduced a definition of avalanche sizes during the 
simulated annealing procedure of the BM and GSA machine, which model memory access.
Both the BM and GSA are regulated by parameters (temperature $T$ and $q_A$ for the 
NSM formalism) which control stochastic fluctuations in network functioning
and model noisy behaviour present in biological synaptic mechanisms.
We conducted simulation experiments with different values of $T$ and $q_A$, and
studied how this causes a change in the behaviour of the two machines,
thus affecting the distribution of avalanche sizes, the network energy loss and measurements
of correlations among energy increments along the path followed during annealing.
In these experiments, we found that for some values of $T$ and $q_A$, the avalanche size
distributions of the GSA machine may be approximately fitted by a $q$-exponential, with 
the corresponding asymptotic power-law behaviour. The BM does not generate this behaviour
and produces smaller avalanche sizes, related to less associative memory functioning,
when compared to GSA.

We conjecture that avalanche sizes defined for our model should be related to the 
time consumed and also to the size of the neuronal 
region which is activated, during access of information in memory.  
This assumption is reasonable, since the transition probabilities~(\ref{eq:BGTransProb}) and~(\ref{eq:TsaTransProb})
which regulate the BM and GSA network functioning should mimic the way the memory apparatus occupies
phase-space, during the retrieval of a mnemenic trace.
We may therefore qualitatively compare the behaviour of the 
distribution of cluster sizes, obtained during fMRI measurements of the
propagation of signals in the brain reported in~\cite{Ta12}, with the distribution of avalanche sizes 
obtained in our simulation experiments. For some values of the $T$ and $q_A$ parameters,
we find a similar power-law like behaviour, spanning similar 
orders of magnitude of avalanche sizes and frequencies, only when using the GSA
machine, which is based on the NSM formalism. The BM did not produce
this pattern in our simulation experiments. This result experimentally corroborates our 
original indication, that the $q$-exponential and asymptotic power-law (scale-free)
behaviour of macroscopic quantities which describe the structure of the complex networks generated by 
the algorithm of our model suggest that they may indeed be better described by the NSM~\cite{Ts96}
formalism, rather than by Boltzmann-Gibbs (BG) statistical mechanics.

We have also considered recent indications that the brains of mammals
are composed of a proportion of inhibitive synapses, which has been measured
to have values between 20 and 30\%~\cite{Ca15}. We therefore measured avalanche size distributions
during memory access, in networks generated by the clustering algorithm of our model with
a proportion of 30\% of inhibitive synapses. In this case, we also found that, for some
$T$ and $q_A$ values, the avalanche size
distributions may be fitted by a $q$-exponential (with the corresponding asymptotic power-law
behaviour), only for the GSA machine. Indeed, the $q$-exponential fit for the case we
analysed seems even better with this proportion of inhibitive synapses than
with 50\% of inhibitive synapses.

Although the fMRI measurements presented in~\cite{Ta12}, capture the
dynamics of the brain in a {\em resting state\/}, the authors argue that  
``... despite the fact that in resting data there are
not explicit inputs, the average BOLD signal around the extracted
points ... still resembles the hemodynamic response function (HRF)
evoked by a stimulus''. This further supports the qualitative comparison of the 
results of our simulation experiments with their fMRI measurements.

We may thus conclude that the computational model studied in this work 
is viable as an illustrative schematic model of some basic mental mechanisms and in fact 
is qualitatively comparable to fMRI data obtained from patients, demonstrating actual brain activity. 
This comparison also reinforces the idea that complex systems such as the brain 
may well be better described by Nonextensive Statistical
Mechanics, which produces the asymptotic power-law behaviour, for
various temperatures and $q_A$ values, while Boltzmann-Gibbs statistical
mechanic does not show such behaviour. In~\cite{Be03}, the authors argue that in the critical
states characterized by power-laws, ``the network may satisfy the competing demands of information 
transmission and network stability''. As demonstrated in earlier work regarding 
this model \cite{We09,We11}, as well as in our simulations, memory retrieval
governed by Nonextensive Statistical Mechanics allows much more associativity in memory functioning
than Boltzmann-Gibbs statistical mechanics, while still achieving the stability necessary for 
information storage (see also~\cite{We16}). This results in a potentially more creative mental structure with 
more capacity for the establishment of metaphors. From the perspective of psychotherapy,
patients with more creativity have a larger possibility of reassociating traumatic and
repressed experiences, and to construct new historical perspectives for their present and 
future, during therapy.
Further investigations along this line of research are important, since they may reveal
more basic structural features of brain and mental functioning.

\begin{acknowledgments}
We are grateful to Constantino Tsallis, Evaldo Curado, Angel R. Plastino and Abbas Edalat 
for fruitful conversations.
This research was developed with grants from the Brazilian National 
Research Council (CNPq), the Rio de Janeiro State Research 
Foundation (FAPERJ) and the Brazilian agency which funds graduate studies (CAPES).

\end{acknowledgments}

\end{document}